\documentclass[aps,prb,reprint,superscriptaddress,floatfix,longbibliography,nofootinbib]{revtex4-2}

\usepackage[colorlinks=true,citecolor=blue,urlcolor=blue,linkcolor=blue]{hyperref}

\usepackage{amsmath,amssymb,mathtools,bm,graphicx,microtype}
\usepackage[T1]{fontenc}
\usepackage{lmodern}
\usepackage{xcolor,booktabs}

\graphicspath{{figures/}}

\newcommand{\ii}{\mathrm{i}}
\newcommand{\ee}{\mathrm{e}}
\newcommand{\eps}{\varepsilon}

\newcommand{\sgn}{\operatorname{sgn}}

\begin{document}

\title{Maximally Chiral Eigenstate from Inverse Thermodynamic Bethe Ansatz}

\author{Tigran A.~Sedrakyan}
\affiliation{Department of Physics, University of Massachusetts, Amherst, Massachusetts 01003, USA}
\affiliation{A.~Alikhanyan National Science Laboratory, Br.~Alikhanian 2, Yerevan 0036, Armenia}

\begin{abstract}
We show that an ordered quantum state can support waves with a finite speed in one direction and exponentially small energies in the other. We prescribe this spectrum for a spin-$1/2$ Heisenberg eigenstate and reconstruct a nonlocal Hamiltonian. The thermodynamic construction selects the ordered, current-carrying state as a ground state. Its magnetic moment remains finite while its response to a weak field has a divergent slope. The entropy per spin vanishes only inversely with a logarithm of temperature, with an exact coefficient containing the golden ratio. Our spin-response calculations reveal a bound pair whose energy can be tuned without changing its intensity. Added exchange and smoother mode selection restore quadratic excitations at the lowest energies. These results connect the design of interactions to directional dynamics and unusual thermodynamics. We propose momentum-resolved spin spectroscopy, weak-field magnetometry, and low-temperature entropy measurements to probe these counterintuitive signatures.
\end{abstract}

\maketitle

\section{Introduction}
\label{sec:intro}

Consider a quantum magnet whose spins form an ordered state. Its wave function fixes the spin correlations, whereas the Hamiltonian sets the energy required to rearrange them. These excitation energies determine how spin motion propagates, how a magnetic field changes the moment, and how many configurations remain accessible on cooling. Can we change these dynamical and thermal properties while preserving the ordered wave function? A particularly sharp possibility is to retain a finite propagation speed in one direction while making collective motion exponentially inexpensive in the other. This would connect spontaneous magnetic order to an unusually large supply of low-energy rearrangements.

Conventional low-energy theories give a useful benchmark. In many one-dimensional quantum fluids, long-wavelength fluctuations form sound-like waves whose energies are proportional to momentum~\cite{HaldaneHarmonicFluid}. Broken spin-rotation symmetry similarly constrains the soft modes of an ordered magnet~\cite{WatanabeMurayama,Leutwyler}. The precise energy law also depends on the spatial structure of the interactions. Replacing a power law by an exponential in wavelength would change more than a propagation speed: it would reorganize the number of states accessible at every low-energy scale.

Excited states make the separation between a wave function and its excitation energies especially useful. In generic interacting systems, most eigenstates in a narrow energy window have thermal local properties~\cite{Deutsch,Srednicki,RigolETH,DAlessio}. Integrable systems possess additional conserved quantities, so their stationary states retain information beyond the energy~\cite{RigolGGE,IlievskiComplete,EsslerFagotti}. The quantum Newton's-cradle experiment demonstrated the physical significance of this distinction: a nearly integrable one-dimensional gas retained a nonthermal momentum distribution through thousands of collisions~\cite{KinoshitaNewtonCradle}. Within an integrable chain, excited eigenstates with completely specified mode occupations can have zero occupation entropy and well-defined low-energy structures of their own~\cite{YangYang,TakahashiBook,VlijmCaux}. They provide a controlled starting point for designing collective motion in an interacting quantum state.

Consider a periodic chain of $L$ spin-$1/2$ moments,
\begin{equation}
 H_{\rm XXX}=J_{\rm X}\sum_{m=1}^{L}
 {\bm S}_m\cdot{\bm S}_{m+1}.
 \label{eq:XXX-intro}
\end{equation}
Here $J_{\rm X}$ is the exchange coupling, ${\bm S}_m$ is the spin operator at site $m$, and ${\bm S}_{L+1}={\bm S}_1$ joins the ends. This isotropic Heisenberg model is also called the XXX chain. Its commuting conserved operators share eigenvectors. Combining them in a new Hamiltonian changes the energies of those vectors without changing their wave functions. An excited Heisenberg eigenstate can thereby become the ground state of a different Hamiltonian~\cite{Tsvelik1990,Frahm1992,SedrakyanLocal,SedrakyanBabujian}, which we call a \emph{selector}. Local selectors already produce ordered, current-carrying excited states~\cite{SedrakyanLocal}. The reference Heisenberg Hamiltonian defines their energy and energy current; the selector defines the excitation spectrum and thermodynamics studied here.

This construction connects to the broader inverse problem of finding interactions compatible with a chosen quantum state. Eigenstate-to-Hamiltonian methods construct families of such interactions, while Hamiltonian-reconstruction methods determine when a single eigenstate fixes a local Hamiltonian~\cite{ChertkovClarkInverse,QiRanardHamiltonian}. Integrability supplies additional design freedom through commuting interactions. Examples include alternating interactions in coupled-chain ladders~\cite{AmbjornLadder,SedrakyanTJ,ArnaudonLadders}, generalized Heisenberg chains with calculable spin and chirality correlations~\cite{MkhitaryanSedrakyanCorrelations}, long-range spin models~\cite{FerrandoCalogero}, and constrained Rydberg-atom chains~\cite{CorcoranRydberg}. These approaches motivate a complementary target: specify not only the wave function, but also the collective energy costs around it.

That target requires the response of the other spins to be included. Adding one excitation changes its scattering with every occupied mode, which readjusts the energy and momentum of the whole state. The same collective readjustment determines the effective velocities in generalized hydrodynamics, a theory that follows the large-scale transport of conserved quantities in integrable systems~\cite{CastroAlvaredoGHD,BertiniGHD}. Here the question is an inverse one: which interaction produces a prescribed excitation energy after this readjustment? Bound groups of reversed spins must also be included, because they can compete with the intended state. Specifying a wave function alone therefore does not settle the dynamical design problem.

We solve this problem for the directional spectrum described above. Let us reverse the usual thermodynamic Bethe ansatz: prescribe the collective excitation energies and reconstruct the energy contributions whose interaction-induced readjustment returns that spectrum. The resulting selector is nonlocal, with spin processes extending across arbitrarily large separations. A positive conserved penalty raises the energies of bound groups of reversed spins. Together, these terms make the chosen occupation a ground state in the infinite-chain thermodynamic description. The same wave function remains an ordered, current-carrying excited eigenstate of the reference Heisenberg model.

A single mechanism links the resulting spectrum to its magnetic and thermal responses. Along the sequence of increasingly soft modes, the excitation energy decreases exponentially, while the density of available modes decreases only as a power. The energy therefore falls much faster than the supply of possible rearrangements is exhausted. This mismatch produces the exponential dependence on inverse momentum, a finite ordered moment with a divergent weak-field susceptibility, and an entropy per spin that vanishes only inversely with a logarithm of temperature. Interacting state counting fixes the coefficients, including a golden-ratio factor in the entropy. Activated energy scales and logarithmic responses are also central to disordered quantum magnets~\cite{Fisher92,Fisher95,IgloiMonthus,VojtaRare}; here they arise from a designed, translation-invariant spectrum rather than spatially rare regions.

Analytic bounds for specified excitation families and exact calculations on short periodic chains connect this thermodynamic construction to an explicit spin operator. The spin response then identifies which excitations a probe creates most strongly. In particular, a conserved interaction shifts a bound-pair spectral line while preserving its intensity as long as the selected ground state is unchanged. Direct observations of freely propagating and bound spin reversals in optical-lattice Heisenberg chains establish the experimental relevance of resolving these distinct processes~\cite{FukuharaBoundMagnons}. For an engineered realization of the selector, our results motivate momentum-resolved spin spectroscopy, weak-field magnetometry, and low-temperature entropy measurements. Added exchange and smoothing restore quadratic excitations at the lowest energies and determine where the activated behavior remains visible. Together, these results turn the design of collective energy laws into linked, testable predictions for dynamics, magnetic order, and thermal state counting.

% BEGIN_INTRO_ROADMAP: the only permitted forward section references.
In Section~\ref{sec:overview}, we present the results and their physical mechanism. Section~\ref{sec:selector} develops the selector and its ordered state. The directional excitation spectrum and spin response of finite periodic chains are discussed in Section~\ref{sec:dynamics}, and its magnetic and thermal consequences are discussed in Section~\ref{sec:soft}. In Section~\ref{sec:exchange}, we examine the effect of added exchange. Section~\ref{sec:locality} examines how the amplitude for moving a reversed spin decreases with distance. Section~\ref{sec:ring} explains the scattering-induced energy shifts in finite periodic chains. Section~\ref{sec:conclusion} brings together the implications and opportunities opened by this construction.

\section{Overview of Results}
\label{sec:overview}

Let us first follow the physical consequences of the construction. The selected state is magnetically ordered, yet spin rearrangements moving in opposite directions have strikingly different energy costs. Left-moving waves retain a finite speed. The energies of elementary right-moving rearrangements become exponentially small as their wavelength grows. The same low-energy modes let a weak field change the magnetization unusually easily and keep many configurations accessible at very low temperature. One mechanism therefore connects magnetic order, directional motion, and thermal state counting.

We study a Heisenberg chain with exchange coupling $J$, supplemented by a conserved chiral interaction. This interaction combines a local three-spin term that favors one handedness with an additional nonlocal conserved charge. The positive coupling constant $\alpha$ multiplies both parts. The nonlocal charge is a Hermitian spin operator that commutes with the Heisenberg Hamiltonian. It includes correlated spin processes across arbitrarily large separations and shifts the energies of common eigenstates without mixing their wave functions. Its role is to adjust the collective excitation costs so that one direction becomes exponentially soft.

We deliberately include a further conserved interaction that raises the energies of bound groups, in which spin reversals preferentially move together. Its positive coupling $\Delta$ sets an extra energy cost: a single bound pair receives $\Delta$, whereas the selected state of unbound reversals receives none. In the thermodynamic description, larger groups receive a cost that grows with their size. We call this added contribution the \emph{bound-state energy penalty}. It is part of the chosen Hamiltonian, rather than an emergent energy scale. We denote the complete interaction by $H_+$, so the model is $H(J)=J\sum_m{\bm S}_m\cdot{\bm S}_{m+1}+H_+$.

The exchange $J$ is distinct from the fixed reference coupling $J_{\rm X}$. The reference Hamiltonian $H_{\rm XXX}$ defines the energy and energy current of the selected wave function; $H(J)$ determines its excitation costs and thermodynamics. Excitation energies, fields, temperatures, and velocities throughout this paper refer to the constructed model unless explicitly labeled XXX.

To describe an infinitely long chain, we use exact equations that count the occupied spin modes and include how their interactions change the energy. In this thermodynamic description, maximal chirality is reached at $J=0$, where the selected ordered state minimizes the energy per site. Its magnetization, reference energy, and reference current are
\begin{align}
 M_0&=0.1921,\qquad
 \frac{e_{\rm XXX}}{J_{\rm X}}=-0.1453,\nonumber\\
 \frac{j_E^{\rm XXX}}{J_{\rm X}^2}&=-0.2440,
 \label{eq:headline-values}
\end{align}
Here $M_0$ is the magnetization per site along a chosen spin direction, $e_{\rm XXX}=\langle H_{\rm XXX}\rangle/L$ is the Heisenberg energy per site, and $j_E^{\rm XXX}$ is its energy-current density. The displayed values are rounded. For positive reference exchange $J_{\rm X}$, the state lies above the antiferromagnetic ground state by a finite energy per site. The finite moment and current identify a nonlocal realization of the rare magnetochiral eigenstates studied in Ref.~\cite{SedrakyanLocal}: ordered, current-carrying excited states of the reference Heisenberg chain.

The zero-field model preserves rotations of all spins together. Taking the infinite-chain limit before removing an infinitesimal field selects a state with a finite magnetization in a chosen direction. This is spontaneous breaking of spin-rotation symmetry, $SU(2)\to U(1)$: only rotations about that direction leave the state unchanged. The current direction is fixed explicitly by the handedness of the interaction.

To understand the excitations, let us describe the state through collective spin-reversal waves, or modes. Each mode extends over the chain; it is not a particular lattice site. A dimensionless continuous label $x$ orders these interacting modes. In the selected state, every allowed unbound mode with $x>0$ is occupied, while every allowed mode with $x<0$ is empty. Occupied means that the wave contributes a spin reversal to the many-spin state; empty means that this allowed wave is unused. The two ranges therefore describe which collective waves are present, not two regions of the physical chain. With each available mode either completely filled or completely empty, the thermodynamic occupation entropy per site is zero, although the physical spins can remain quantum entangled. Emptying an occupied mode creates a hole and increases the spin along the chosen direction by one. Filling an empty mode adds a particle and decreases it by one. A particle and a hole together preserve the total spin. Their physical energies and momenta include the readjustment of the other occupied modes.

Let us write $q>0$ for the magnitude of a small physical momentum. A spin-preserving excitation moving to the left has energy $E_{\rm F}(-q)$,
\begin{equation}
 E_{\rm F}(-q)=|v_L|q+o(q),\qquad
 v_L\simeq-2.2132\alpha ,
 \label{eq:headline-linear}
\end{equation}
where $v_L$ is its velocity and $o(q)$ denotes terms that vanish faster than $q$. This linear energy law describes a wave with a finite long-wavelength propagation speed. The elementary right-moving excitations have a qualitatively different relation between energy and momentum. Writing $\Omega=\pi^2\alpha/2$, their hole and particle energies, denoted $\omega_+(q)$ and $\omega_-(q)$, obey
\begin{align}
 \omega_+(q)&\simeq\Omega\ee^{\kappa_0}
 \left(\frac{2M_0}{q}\right)^{1/2}\ee^{-\pi M_0/q},
 \nonumber\\
 \omega_-(q)&\simeq\Omega\ee^{-\kappa_0}
 \left(\frac{q}{4M_0}\right)^{1/2}\ee^{-2\pi M_0/q},
 \label{eq:headline-essential}
\end{align}
as $q\to0^+$, with the dimensionless constant $\kappa_0\simeq1.0373$ fixed by the exact solution. The subscripts $+$ and $-$ label the increase and decrease of spin. The factors multiplying the exponentials determine the precise asymptotic energies. The exponentials establish the striking physical contrast: both energies vanish faster than every positive power of $q$. Their slopes, which give the group velocities, also vanish faster than any power. Doubling the wavelength therefore reduces the excitation energy by an exponentially large factor, while the opposite direction retains the finite velocity in Eq.~\eqref{eq:headline-linear}. We call the exponential law an \emph{activated dispersion} and this limiting directional contrast \emph{maximal chirality}.

A spin-preserving right-moving rearrangement combines one hole and one particle. We denote their physical momentum changes by $q_h>0$ and $q_p>0$, with $q_h+q_p=q$. The smallest energy within this single-pair family is
\begin{align}
 \omega_0(q)&=\min_{q_h+q_p=q}
 \big[\omega_+(q_h)+\omega_-(q_p)\big]\nonumber\\
 &\simeq\Omega\ee^{-\kappa_0/3}
 \left(\frac32\right)^{5/6}
 \left(\frac{q}{M_0}\right)^{1/6}
 \ee^{-3\pi M_0/q} .
 \label{eq:pair-threshold}
\end{align}
The minimization has a simple physical meaning. At the optimal division of momentum, moving a little momentum from one constituent to the other leaves the energy unchanged to first order. Their energy slopes, and hence their group velocities, are equal. The division approaches $q_h=q/3$ and $q_p=2q/3$. A logarithmic correction to this division fixes the displayed $q^{1/6}$ factor.

The single-pair result also explains why elementary and many-particle excitations have different energy limits. When many pairs share a fixed momentum, each pair carries less momentum and costs exponentially less energy. Adding their costs therefore allows the total energy to approach zero. We take the chain length to infinity at fixed pair number before increasing the number of pairs. Because lattice momentum is defined modulo $2\pi$, this thermodynamic construction reaches arbitrarily small energies at either momentum sign. Maximal chirality describes the sharply different elementary branches within that larger spectrum.

To see where the exponential laws come from, let us follow the modes as the magnitude of their label $x$ grows. Along the sequence of softer modes, energy decreases exponentially with $|x|$, while the number of modes per unit $x$ decreases only as $1/x^2$. Momentum counts the modes remaining before the occupation boundary, so it decreases as $1/|x|$. Eliminating $x$ gives an exponential in inverse momentum. The corrections to this counting fix the momentum-dependent factors in Eqs.~\eqref{eq:headline-essential} and \eqref{eq:pair-threshold}. In the ordinary antiferromagnetic Heisenberg ground state, energy and momentum have matching exponential tails, which give a linear dispersion~\cite{GriffithsMagnetization,Takahashi1971,EggertAffleckTakahashi}.

A weak oscillating magnetic field provides a way to probe these spin motions. Peaks in the response reveal the energies needed to excite them, while the area under each peak measures the strength of the signal. Our calculations on short periodic chains show a striking contrast for a probe that lowers the spin projection along the ordered moment. In one momentum direction, the strongest signal comes from an unbound spin reversal. In the opposite direction, it comes from two reversed spins bound together.

The bound pair also offers a way to separate the energy of an excitation from its visibility in a measurement. The additional interaction introduced above raises the pair's energy relative to the unbound ground state. Increasing its strength therefore moves the peak to higher frequency. Yet the area under the peak stays fixed: the interaction preserves the wave functions, so the probe excites the pair just as strongly. As long as the same ground state remains selected, the energy needed to create the pair can thus be tuned independently of the strength of its signal.

The probe orientation provides another useful distinction. A field parallel to the ordered moment couples to the spin component along that direction and measures the \emph{longitudinal} response. The \emph{transverse} response concerns the perpendicular spin components; a circularly polarized probe of the appropriate handedness lowers the total spin projection by one. For either response, a line's weight is the squared matrix element of the corresponding spin operator between the initial and final states. This quantity determines the transition strength after the probe amplitude is factored out.

Let $S$ be the total spin of the ground state, with maximal projection along the chosen axis. A longitudinal excitation with the same total spin $S$ has a partner obtained by lowering its spin projection by one. The partner has the same total spin and excitation energy, but is reached by the transverse spin-lowering probe. Spin-rotation symmetry fixes its transverse line weight to $2/S$ times the longitudinal weight of the original excitation. These are \emph{absolute} line weights, before each is divided by the total weight in its response channel. This identifies the same collective rearrangement in two probe orientations.

A weak field turns these inexpensive rearrangements into a large magnetic response. Let us choose the $z$ axis along the ordered moment and write $S^z_{\rm tot}=\sum_m S_m^z$ for the total spin projection. A uniform field $h>0$ adds $-hS^z_{\rm tot}$ to the selector. It favors emptying occupied spin-reversal modes and therefore increases $M(h)=\langle S^z_{\rm tot}\rangle/L$. The removal energies decrease exponentially along the mode sequence, while the number of modes decreases only as a power. As $h\to0^+$, the resulting magnetization is
\begin{equation}
 \begin{aligned}
 M(h)&=M_0+\frac{M_0}{2\ln(\sqrt{2\pi/\ee}\,\Omega/h)}\\
 &\quad+O\!\left(\frac{\ln\ln(\sqrt{2\pi/\ee}\,\Omega/h)}{[\ln(\sqrt{2\pi/\ee}\,\Omega/h)]^2}\right),
 \end{aligned}
 \label{eq:headline-field}
\end{equation}
The $O(\cdots)$ term specifies the scale of the omitted corrections. The fixed factor $\sqrt{2\pi/\ee}$ accounts for the readjustment of the other occupied modes. The change in magnetization disappears only inversely with a logarithm of field. Correspondingly, the longitudinal susceptibility $\chi_\parallel(h)=dM/dh$ diverges as
\begin{equation}
 \chi_\parallel(h)\simeq
 \frac{M_0}{2h[\ln(\sqrt{2\pi/\ee}\,\Omega/h)]^2} .
 \label{eq:headline-susceptibility}
\end{equation}
The moment approaches the finite value $M_0$, but its one-sided slope becomes infinite. These statements describe two different properties of the same state: the alignment already present and the energy required to increase it. The activated spectrum makes that additional alignment exceptionally inexpensive.

\begin{figure*}[t]
 \centering
 \includegraphics[width=\textwidth]{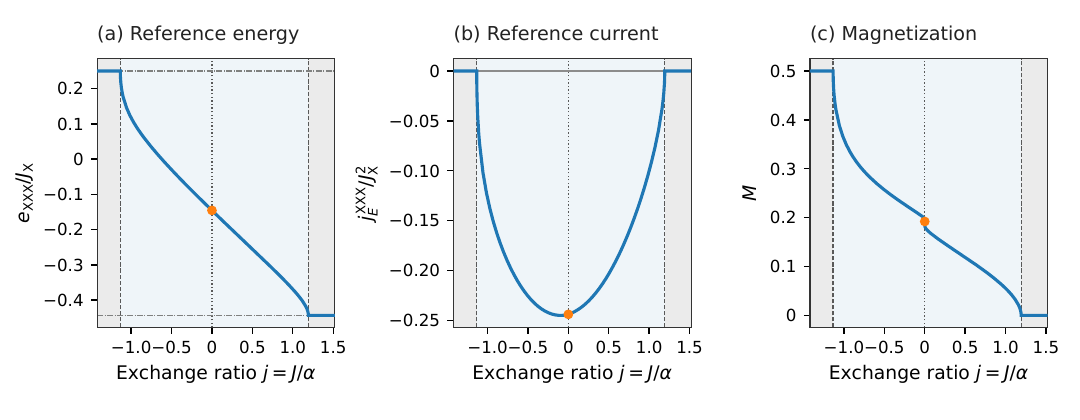}
 \caption{An ordered state with a current between familiar magnetic limits. The horizontal variable $j=J/\alpha$ is the added exchange in units of the selector coupling. The three panels show (a) the scaled Heisenberg reference energy, (b) the scaled Heisenberg energy current, and (c) the magnetization per site. Orange points mark the original model at $j=0$. The outer dashed lines delimit the partially magnetized states carrying a current; beyond them the selected wave functions are the fully aligned and antiferromagnetic reference states. The central dotted line marks the activated point. Arbitrarily weak added exchange restores quadratic excitations at its very soft boundary, while the magnetization remains continuous with a divergent slope.}
 \label{fig:exchange}
\end{figure*}

Temperature probes how many such configurations can be occupied. For a chain of $L$ sites, let $S_{\rm th}(L,T)$ be the thermal entropy at temperature $T$. The entropy density is
\begin{equation}
 s(T)=\lim_{L\to\infty}\frac{S_{\rm th}(L,T)}{L} .
 \label{eq:entropy-density-definition}
\end{equation}
At fixed nonzero temperature, the leading total entropy is extensive: $S_{\rm th}(L,T)=Ls(T)+o(L)$. In one dimension, an ordinary linear branch gives $s\propto T$, while a quadratic branch gives $s\propto\sqrt T$. These powers reflect the momentum interval with excitation energies below the thermal energy. A nonzero excitation gap instead suppresses entropy exponentially on cooling.

Here that interval shrinks much more slowly, as $1/\ln(\Omega/T)$. The number of thermally active modes is therefore of order $L/\ln(\Omega/T)$, and the zero-field entropy density obeys
\begin{equation}
 s(T)\simeq\frac{2M_0\ln\varphi}{\ln(\Omega/T)},
 \qquad \varphi=\frac{1+\sqrt5}{2},
 \label{eq:headline-entropy}
\end{equation}
with $T\to0$ taken after $L\to\infty$. The golden ratio $\varphi$ appears because the modes are coupled: occupying one changes the number of states available to the others. The exact thermal counting includes this interaction and fixes the coefficient. Thus the entropy density still vanishes, but much more slowly than a power of temperature. This behavior comes from an extensive set of low-energy configurations. A single spin multiplet, whose degenerate states are related by global spin rotations, contributes at most $\ln(L+1)$ to the total entropy and cannot account for it. The thermal result counts configurations in the spin-rotation-invariant, zero-field Gibbs ensemble.

The same mechanism predicts how switching on the Heisenberg exchange changes the unusual spectrum. Figure~\ref{fig:exchange} follows $H(J)$ away from the maximally chiral point $J=0$. The horizontal coordinate $j=J/\alpha$ measures the exchange relative to the chiral coupling. The three panels follow the reference energy, reference current, and magnetization of the ground state selected at each $j$. Negative exchange favors parallel neighboring spins, while positive exchange favors antiparallel correlations. Between the fully aligned and antiferromagnetic limits lies a family of partially magnetized states carrying a reference current. The orange points at $j=0$ identify the original state and the values in Eq.~\eqref{eq:headline-values}.

Magnetic order and current persist for nearby nonzero $J$. An arbitrarily weak nonzero exchange eventually exceeds the exponentially small selector energy at sufficiently small momentum. For either sign of $j$, it restores quadratic excitations at the very soft boundary. The magnetization remains continuous. As $j\to0$, it obeys
\begin{equation}
 M(j)=M_0-\sgn(j)\frac{M_0}{2\ln(1/|j|)}
 +O\!\left(\frac{\ln\ln(1/|j|)}{\ln^2(1/|j|)}\right).
 \label{eq:M-main}
\end{equation}
Here $\sgn(j)$ is the sign of $j$. The slope of $M(j)$ diverges. Figure~\ref{fig:exchange}(c) therefore connects a continuous ordered moment to a singular change of the long-wavelength energy law. The central model is an isolated singular point along this exchange deformation. At the outer dashed boundaries, new low-energy modes appear with energy proportional to momentum squared. Related logarithmic laws occur in disordered magnets~\cite{Fisher92,Fisher95,IgloiMonthus,VojtaRare}; here they arise from the interacting spectrum of a translation-invariant chain with no random couplings.

The construction also identifies the spatial structure that supports these excitations. The selector is nonlocal: its interactions can connect spins at arbitrarily large separations. A directly calculable measure in the infinite-chain additive description is $t(r)$, the matrix element for moving one reversed spin by $r$ lattice sites in an otherwise fully aligned chain. At large $r$,
\begin{equation}
 \frac{|t(r)|}{\alpha}\propto
 r^{-3/4}\ee^{-\sqrt{2\pi r}}.
 \label{eq:stretched}
\end{equation}
This decay is faster than every power of distance and slower than an exponential. An exact asymptotic evaluation fixes its amplitude and first correction. One- and two-spin matrix elements further reveal the spatial reach of the selector and its correlated interactions. Making the selector's sharp mode-selection rule more gradual restores a quadratic branch at the smallest momenta, while an activated crossover survives at larger momenta. The exchange and smoothing tests both expose the cancellation that sustains the exponential energy law.

Finite periodic chains provide a further test. Scattering shifts the allowed momenta and adds an energy correction that can exceed the exponentially small infinite-chain cost. We isolate this correction by comparing specified pairs of eigenstates. Independent ground-state searches then track how a weak field selects the ordered state. Together, these calculations resolve how the predicted soft motion appears in systems of finite size.

The common route is to prescribe the interacting excitation energies and reconstruct a Hamiltonian that realizes them. A continuous family of commuting conserved quantities supplies the required freedom~\cite{IlievskiStringCharge,IlievskiQuasilocal,IlievskiLandscape}. Earlier selectors with nearby-spin interactions establish ordered, current-carrying excited states with power-law excitation energies~\cite{SedrakyanLocal,WeiChiral}. The present construction extends that control to an activated energy law. Exact integral equations determine its magnetic and thermal consequences, while energy bounds ensure that bound groups of reversed spins remain costly. Exchange and smoothing then provide direct ways to tune the unusually soft motion.

We connect this thermodynamic description to the spin operator through complementary checks. For eigenstates with real mode parameters satisfying the stated occupation, spin-density, and spacing conditions, a finite-size theorem bounds the difference between microscopic and assigned energies by a stretched exponential in chain length. Exact calculations test bound-spin excitations and the states selected by a field. A condition on thermally weighted energy corrections specifies when the spin operator has the same free energy. These results make the connection between the designed spectrum and its microscopic realization precise.

\section{How the state and its spectrum are constructed}
\label{sec:selector}
\label{sec:maximal}

Let us first change the energy of a Heisenberg eigenstate while preserving its wave function. Commuting conserved operators share eigenvectors, so their coefficients determine the energies assigned to those vectors. Our starting point is
\begin{align}
 Q_2&=\sum_{m=1}^{L}{\bm S}_m\cdot{\bm S}_{m+1},
 \label{eq:Q2-main}\\
 Q_3&=\sum_{m=1}^{L}{\bm S}_m\cdot
 ({\bm S}_{m+1}\times{\bm S}_{m+2}).
 \label{eq:Q3-main}
\end{align}
The charge $Q_2$ measures neighboring-spin correlations, while $Q_3$ measures the handedness of three spins. Both commute with $H_{\rm XXX}=J_{\rm X}Q_2$, whose total energy current is $J_E^{\rm XXX}=J_{\rm X}^2Q_3$~\cite{Subrahmanyam,ZotosCurrent,GrabowskiMathieu}. We can therefore favor a current-carrying eigenstate by lowering its energy through a suitable combination of these charges. The reference coupling $J_{\rm X}$ is distinct from the added exchange $J$. State preparation changes eigenstate populations; varying commuting charge coefficients alone preserves them.

Changing the coefficients also changes the thermal equilibrium associated with the selector. At temperature $T$, an eigenstate with selector energy $E$ has a probability proportional to $\exp(-E/T)$. Its energy is a weighted sum of its conserved-charge values. States whose energies increase relative to the others receive less equilibrium weight, while their wave functions remain unchanged. This compares equilibrium ensembles of different Hamiltonians; a closed-system change of commuting coefficients alone still preserves the initial populations. Related exponential weighting is used to study rare currents in stochastic lattice gases~\cite{BertiniCurrentFluctuations,BertiniMFT}. There one reweights entire time histories according to the current transported during an observation. Here one weights stationary quantum eigenstates according to their energies and conserved quantities.

To describe the selected state, let us begin with one reversed spin spread coherently over the chain with momentum $p$:
\begin{equation}
 |p\rangle=\frac1{\sqrt L}\sum_m\ee^{\ii pm}
 S_m^-|{\uparrow\cdots\uparrow}\rangle,
 \label{eq:one-spin-wave}
\end{equation}
Here $S_m^-=S_m^x-\ii S_m^y$ and $\ii^2=-1$. This spin-reversal wave is a \emph{magnon}. We denote total momentum by $P$ and use the translation operator $\mathcal T=\ee^{-\ii P}$. Several waves scatter, and the Bethe ansatz constructs their joint wave function from the phases acquired in their encounters.

It is convenient to label each wave by a dimensionless variable $x$, called its \emph{rapidity}. Its momentum before the other occupied waves readjust is
\begin{equation}
 p(x)=2\arctan x-\pi,\qquad
 p'(x)=\frac{2}{1+x^2}>0.
 \label{eq:momentum-convention}
\end{equation}
The allowed values $x_j$ are the \emph{Bethe roots}. Their number $N$ counts reversed spins, so $S^z_{\rm tot}=L/2-N$. This count refers to the fully aligned reference in Eq.~\eqref{eq:one-spin-wave}: each spin configuration has $N$ down spins, and the eigenstate is a coherent sum of these configurations. The reversals are therefore spread over the chain.

Let us distinguish two ways that these reversals can move. In an \emph{unbound} state, they can move apart without the wave amplitude becoming exponentially small as their separation grows. They still scatter and change one another's phases. In a bound pair, the amplitude instead decreases exponentially with the separation: the two reversals preferentially travel together. The pair as a whole can travel along the chain. Binding describes this dependence on separation; two neighboring down spins in a single configuration do not by themselves establish a bound state. Real Bethe roots label unbound spin-reversal waves. A correlated group of $n>1$ roots with nonreal members, called an $n$-string, describes $n$ reversals bound together.

Our selected state contains a finite density of unbound waves and fills their modes with $x>0$. Adding a particle fills an empty mode, while creating a hole empties an occupied one. The chain still contains $L$ physical spins. These are excitations of the partially magnetized selected state. In either case, scattering shifts the other occupied modes. The physical excitation momentum $\Delta P$ includes their collective readjustment.

In the zero-temperature thermodynamic description, the \emph{dressed energy} $\eps(x)$ measures the change in energy on increasing the occupation near $x$, including the scattering-induced shifts of other modes. Its sign tells us which modes the ground state favors. A negative value means that occupation lowers the energy, so that mode is filled; removing it costs the positive energy $-\eps(x)$. A positive value means that adding a particle costs $\eps(x)$, so the mode remains empty. Thus a negative dressed energy describes a favorable occupation, not a negative excitation cost above the ground state. The boundaries between occupied and empty modes occur where $\eps$ vanishes.

The \emph{bare energy} $w_1(x)$ is the contribution assigned directly to an unbound mode by the Hamiltonian, before including the changes in the other modes. The thermodynamic Bethe ansatz (TBA) relates this contribution to the dressed energy~\cite{YangYang,ZamolodchikovTBA,TakahashiBook}. Let us reverse that relation: prescribe the dressed energy $\eps_0(x)$, then determine the bare energy $w_1(x)$ that produces it. For real roots we choose
\begin{equation}
 \eps_0(x)=2\pi\alpha s'(x)
 =-\frac{\pi^2\alpha}{4}\frac{\sinh(\pi x/2)}{\cosh^2(\pi x/2)},
 \label{eq:eps0-main}
\end{equation}
Here $s(x)=1/[4\cosh(\pi x/2)]$, and the subscript $0$ denotes the selected zero-field state. This choice has $\eps_0(x)<0$ for $x>0$ and $\eps_0(x)>0$ for $x<0$, giving precisely the occupation pattern described above. Its magnitude becomes exponentially small at large $|x|$: removing a distant occupied mode or adding a distant empty one therefore costs exponentially little energy. These are the activated excitations introduced in Section~\ref{sec:overview}.

Let us now turn the prescribed excitation energies into a spin Hamiltonian. Adding one spin-reversal wave shifts the other occupied waves through scattering. The bare energy $w_1$ must compensate this readjustment so that the resulting mode energy is the chosen $\eps_0$. We therefore add a conserved observable to the local charge $Q_3$. Here a \emph{charge} means an observable whose value remains constant under Heisenberg time evolution. The charges used in our construction share eigenvectors, so their linear combination assigns new energies to the same wave functions. We denote the required additional, dimensionless observable by $\mathcal Q_{\rm nl}$.

To construct this observable, let us introduce an auxiliary spin one with spin operator ${\bm S}_a$. Its three states, with projections $-1,0,1$, span a space $V_a$. The auxiliary spin is a mathematical device, not an added site in the physical spin-$1/2$ chain. A dimensionless parameter $u$ labels the operator family.

For site $m$, the local matrix $L_{2,m}(u)$ acts on the auxiliary spin and the physical spin ${\bm S}_m$. The subscript $2$ denotes twice the auxiliary spin.
In the auxiliary basis with spin projections $+1,0,-1$, the local factor contains only the on-site operators $S_m^z$ and $S_m^\pm=S_m^x\pm\ii S_m^y$:
\begin{equation*}
 L_{2,m}(u)=
 \begin{pmatrix}
 u+2\ii S_m^z&\sqrt2\,\ii S_m^-&0\\
 \sqrt2\,\ii S_m^+&u&\sqrt2\,\ii S_m^-\\
 0&\sqrt2\,\ii S_m^+&u-2\ii S_m^z
 \end{pmatrix},
\end{equation*}
The transfer matrix is
\begin{equation*}
 T_2(u)=\operatorname{tr}_{V_a}
       [L_{2,L}(u)\cdots L_{2,1}(u)] .
\end{equation*}
Scalar entries multiply the identity on site $m$.
The ordered product combines the auxiliary spin's successive encounters with sites $1$ through $L$, summing over all its intermediate spin projections. The trace $\operatorname{tr}_{V_a}$ sums the diagonal auxiliary matrix elements: for each starting projection, it retains the contribution with the same final projection, then adds the three results. Removing the auxiliary index leaves an operator on the original $2^L$-dimensional spin-chain Hilbert space, the \emph{transfer matrix}~\cite{FaddeevABA,KorepinBook}.

The exact scattering relations make these transfer matrices commute with one another and with the Heisenberg Hamiltonian. To fix the nonlocal interaction, including its normalization, let us define
\begin{equation*}
 \mathcal F_2(x)=\frac{T_2((x+\ii)/2)}{[(x+3\ii)/2]^L},
 \qquad
 D_2(x)=\mathcal F_2(x)^{-1}\partial_x\mathcal F_2(x).
\end{equation*}
The real, dimensionless parameter $x$ labels this family. The denominator removes the leading scalar factor, while the derivative measures how the remaining spin operator changes with $x$. The Hermitian combination $(D_2-D_2^\dagger)/(4\pi\ii)$ is the conserved charge $X_2(x)$; the dagger denotes the Hermitian adjoint. An isolated zero is treated by the symmetric limiting prescription in the operator integral.
The same construction with an auxiliary spin $1/2$ gives the family $T_1(u)$ that generates $Q_2$ and $Q_3$. The spin-one family $T_2(u)$ supplies the additional freedom required here.

Its usefulness follows from a direct match to scattering. Let us label the Hermitian charge family by real $x$, using $u=(x+\ii)/2$. In the thermodynamic description, a real Bethe root at $y$ contributes $a_2(x-y)$, where $a_n(x)=(n/\pi)/(x^2+n^2)$ for positive integer $n$. Here $x$ selects a conserved operator, while $y$ labels an occupied mode. The same function $a_2$ determines the readjustment of the other unbound modes. A weighted sum of these conserved operators can therefore supply exactly the correction required by the inverse TBA. Combining them over negative $x$, with real weights fixed by $\eps_0$, defines $\mathcal Q_{\rm nl}$. Together with $Q_3$, it yields $w_1$ and hence the desired collective spectrum. The ordered products contain correlated spin processes spanning arbitrarily large separations. This spatial reach is the meaning of ``nonlocal.''

Let us next favor this state of unbound spin-reversal waves over states containing bound groups. The construction so far assigns a bound group zero dressed energy: after the occupied waves readjust, adding the group costs no energy in the thermodynamic description. Such groups would therefore introduce additional states at the same minimum energy. We remove this degeneracy by assigning a positive energy to bound groups. The unbound waves, whose reversals can separate along the chain, keep their prescribed mode energies.

This extra term must also preserve the freedom to rotate all spins together. For a state with total spin $S$, the $2S+1$ possible spin projections form a \emph{multiplet}. Let us describe its internal structure using the member with the largest projection, $S^z_{\rm tot}=S$, called the highest-weight state. We assign the same penalty to every member of that multiplet. The penalty then depends on the internal pattern of bound reversals and treats all spin orientations equally.
Let $\lambda$ label a common Heisenberg eigenmultiplet, let $x_{\lambda j}$ be the finite Bethe roots of its highest-weight member, and let $\Pi_\lambda$ project onto all its spin orientations. The corresponding operator is
\begin{equation}
 \begin{aligned}
 c_\lambda&=\tfrac12\,\#\{j:\operatorname{Im}x_{\lambda j}\ne0\},\\
 R_C&=\sum_\lambda c_\lambda\Pi_\lambda,\qquad
 H_{\rm reg}=\Delta R_C,\quad\Delta>0.
 \end{aligned}
 \label{eq:additive-regulator-main}
\end{equation}
With these definitions, the complete interaction in $H(J)$ is
\begin{equation}
 \begin{aligned}
 H_+={}&\alpha\sum_{m=1}^{L}{\bm S}_m\cdot
          ({\bm S}_{m+1}\times{\bm S}_{m+2})+\Delta R_C\\
 &-\frac{\alpha}{2\ii}\int_{-\infty}^{0}s'(x)
                 [D_2(x)-D_2(x)^\dagger] \,dx .
 \end{aligned}
 \label{eq:full-selector-main}
\end{equation}
The three-spin term fixes a handedness. The integral is the nonlocal contribution $\alpha\mathcal Q_{\rm nl}$ that adjusts the mode energies; its weight $s'(x)$ was defined in Eq.~\eqref{eq:eps0-main}. The term proportional to $\Delta$ assigns the chosen extra energy to bound-root multiplets. Every transfer-matrix entry is written in physical spin operators above. We  construct each $\Pi_\lambda$ explicitly from products of $S_m^z$ and $S_m^\pm$ and derives the full expression from the rational scattering matrix. This gives a finite-chain spin Hamiltonian, while the energy-minimization argument below concerns its additive thermodynamic description.
The symbol $\#$ counts the finite Bethe roots whose imaginary parts are nonzero. They occur in conjugate pairs, with equal real parts and opposite imaginary parts. The factor $1/2$ counts each pair once. Thus $R_C=0$ for a state containing only real roots, while each nonreal pair contributes exactly $\Delta$ through $H_{\rm reg}$. The other charges supply the rest of the state's energy. This rule defines an operator by specifying the number that multiplies each common eigenstate. The numbers are real, so $R_C$ is Hermitian; leaving these eigenstates in place makes it commute with the Heisenberg Hamiltonian. The assignment also preserves translations and spin rotations. Since the highest-weight state has at most $L/2$ finite roots, $R_C$ is an integer between zero and $L/4$.

The counting has a simple physical interpretation in the ideal-string description. A group of two or three bound reversals contains one conjugate pair and costs $\Delta$; a group of four or five contains two pairs and costs $2\Delta$. In general, an ideal $n$-string contributes $\lfloor n/2\rfloor$, where $\lfloor\cdot\rfloor$ denotes the integer part. On finite chains, deviations from ideal strings and pairs collapsing onto the real axis can change this count~\cite{HagemansCauxStrings}. When their total discrepancy grows more slowly than $L$, written $o(L)$, its contribution per site vanishes as the chain grows. The penalty density then approaches $\Delta\sum_{n\ge2}\lfloor n/2\rfloor\int\rho_n(x)\,dx$. Here $\rho_n$ counts occupied $n$-string modes per site and unit rapidity. Each term multiplies the cost of one bound group by the number of such groups per site. Their dressed energies are then $\eps_{n\ge2}=\lfloor n/2\rfloor\Delta$, while $\eps_1=\eps_0$. Bound groups acquire positive excitation costs, and the selected unbound modes retain the exponentially soft spectrum.

Ground-state selection is transparent in the thermodynamic description. The energy, written as a sum over modes, differs from the target by
\begin{equation}
 E-E_0=L\sum_n\int dx\,
 \left[\eps_n^+\rho_n+|\eps_n^-|\rho_n^{\rm h}\right],
 \label{eq:linear-identity}
\end{equation}
Here $E_0$ is the target energy, $\rho_n^{\rm h}$ counts empty modes, and $\eps^+=\max(\eps,0)$, $\eps^-=\min(\eps,0)$. Every term is nonnegative. Occupying a positive-energy mode or emptying a negative-energy mode costs energy. Filling precisely the real-root modes with $x>0$ makes all these costs vanish. This establishes the global thermodynamic minimum, with the collective effects of scattering already included.

The exact spin operator on a finite periodic chain supplies a direct check. Its energy converges to the additive assignment faster than every power of system size for real-root sequences with nonzero spin density, at most a fixed number of negative roots, and the stated upper bound on the density of positive roots. Exact states containing bound groups reveal another mechanism: contributions in the transfer-matrix expression cancel a finite term present in an isolated contribution. This brings the full energy close to the ideal-string value in the computed families and connects the designed spectrum to the spin operator.

Let us now determine the selected state's magnetization and current. The XXX scattering phases fix the number of occupied modes per site and unit rapidity:
\begin{equation}
 \rho_0(x)+\int_0^\infty a_2(x-y)\rho_0(y)\,dy=a_1(x),
 \qquad x>0 .
 \label{eq:rho0-main}
\end{equation}
Solving this equation on $x>0$ by a Fourier-factorization method~\cite{TakahashiBook,MkhitaryanSedrakyanZigzag} gives the observables in Eq.~\eqref{eq:headline-values}, including
\begin{equation}
 \frac{\langle Q_3\rangle}{L}=-\pi\rho_0(0)^2,
 \qquad \rho_0(0)\simeq0.2787 .
 \label{eq:chirality-density}
\end{equation}
The occupied fraction is $N_0=\int_0^\infty\rho_0(x)\,dx=0.3079$, giving $S/L\to M_0=1/2-N_0$. Removing a selecting field after the infinite-chain limit chooses a magnetization direction and breaks spin rotations from $SU(2)$ to $U(1)$. The chiral coupling fixes the current's handedness.

Every mode is filled or empty, so the occupation-counting entropy, called the Yang--Yang entropy, is zero. The wave function carries reference XXX current density $-0.2440J_{\rm X}^2$; evaluating the additive generalized-current expression~\cite{BorsiCurrents,PozsgayCurrentAlgebra} for the selector gives zero flux.

\begin{figure*}[t]
 \centering
 \includegraphics[width=0.94\textwidth]{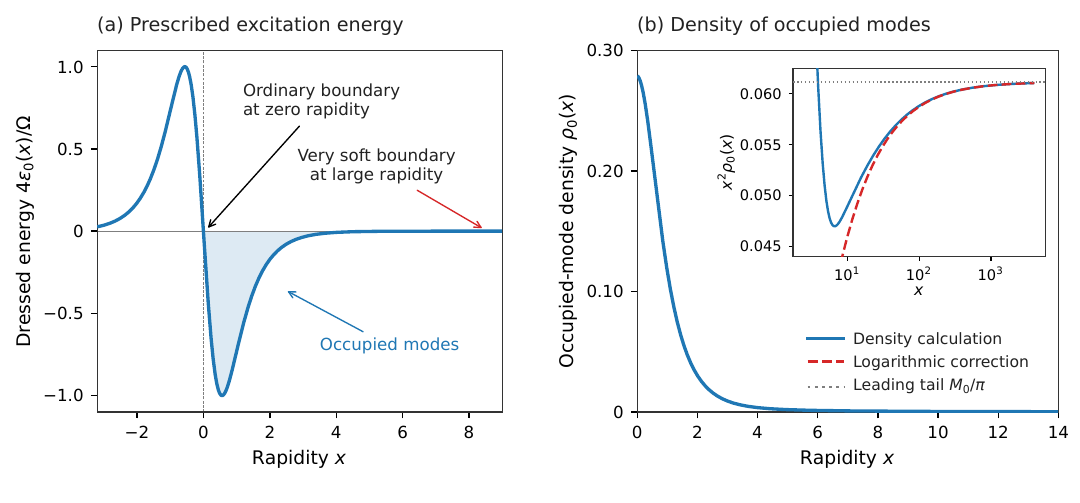}
 \caption{The selected state and its two occupation boundaries. (a) The target dressed energy $\eps_0(x)$, plotted as $4\eps_0(x)/\Omega$ with $\Omega=\pi^2\alpha/2$, is negative on the occupied side. The dimensionless rapidity $x$ labels interacting spin-reversal modes, and $\alpha$ fixes the selector's energy scale. The ordinary boundary lies at $x=0$. The very soft boundary is approached at rapidity infinity. (b) Occupied-mode density $\rho_0(x)$, per site and per unit rapidity. The inset shows $x^2\rho_0(x)$ approaching $M_0/\pi$, where $M_0\simeq0.1921$ is the magnetization per site. The leading inverse-square tail alone gives the gray horizontal line. Including the first logarithmic and constant corrections gives the red dashed curve and explains the slow approach to that limit.}
 \label{fig:maximal-state}
\end{figure*}

Figure~\ref{fig:maximal-state} shows why this ordered state can support unusually inexpensive rearrangements. In panel (a), negative mode energies favor occupation on the $x>0$ side. The curve crosses zero with a finite slope at $x=0$, but approaches zero exponentially at large $|x|$. Panel (b) counts the occupied modes: the area under $\rho_0(x)$ is the number of reversed spins per site. Its tail looks small on the main plot. The inset multiplies the density by $x^2$ and reveals its approach to the nonzero value $M_0/\pi$, establishing $\rho_0(x)\sim M_0/(\pi x^2)$. The red dashed curve includes the first logarithmic correction and explains the slow approach to this limit. Thus the number of available rearrangements decreases only as a power along the tail, while their energies decrease exponentially. Many rearrangements of the ordered state therefore require exceptionally little energy.

\section{Directional excitation energies}
\label{sec:dynamics}

\begin{figure*}[t]
 \centering
 \includegraphics[width=\textwidth]{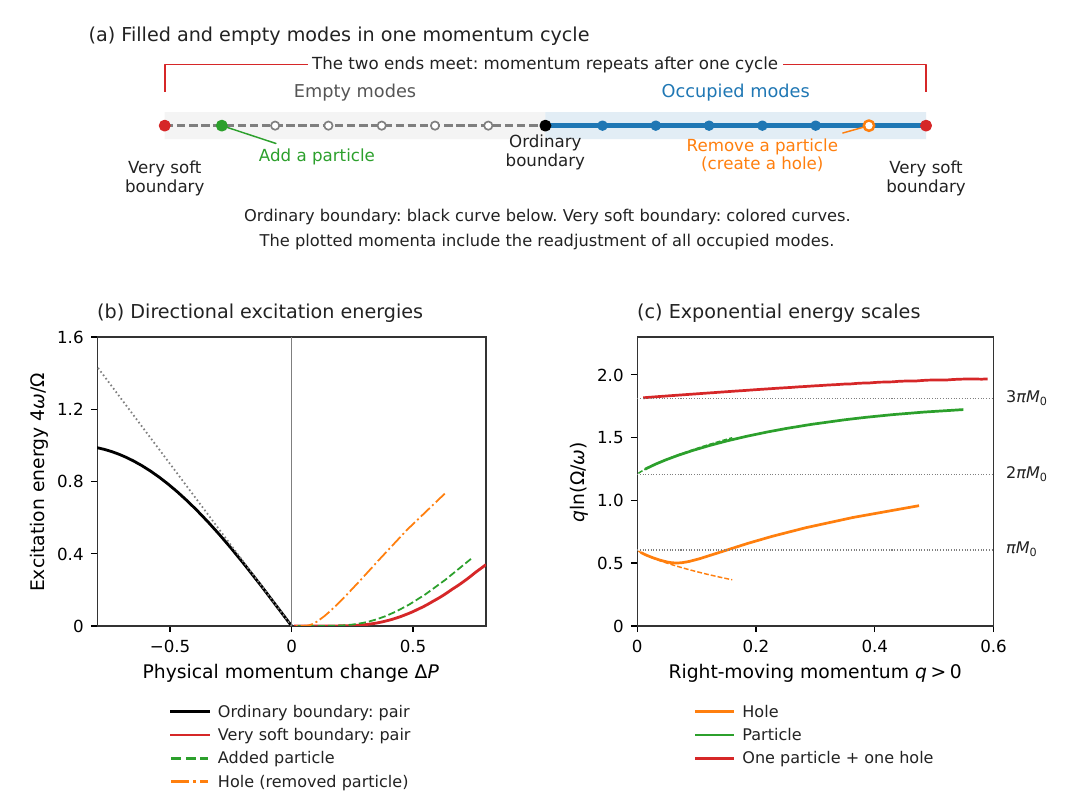}
 \caption{From occupied spin modes to directional excitation energies. (a) Each dot labels a delocalized spin-reversal mode. Open gray dots are empty and blue dots are occupied; their spacing is schematic. Moving from a blue mode to an empty one adds a particle and creates a hole, preserving total spin. The middle of the strip is the ordinary boundary at $x=0$. Its two ends join because momentum repeats after $2\pi$; they form the very soft boundary at $x\to\pm\infty$. (b) Excitation energy $4\omega/\Omega$ versus physical momentum change $\Delta P$, including the readjustment of the other modes. Black shows the minimum energy of one particle--hole pair near the ordinary boundary and the dotted line its finite-slope limit. At positive momentum, green is an added particle, orange a hole, and red the minimum for one particle--hole pair near the very soft boundary. The particle lowers total spin by one, the hole raises it by one, and their pair preserves it. (c) The same positive-momentum energies transformed to expose their exponential dependence. As $q=\Delta P\to0^+$, $q\ln(\Omega/\omega)$ approaches $\pi M_0$, $2\pi M_0$, and $3\pi M_0$ for the hole, particle, and pair. A larger limit means a faster fall of the energy. Dashed segments include the particle and hole prefactors. Here $\Omega=\pi^2\alpha/2$ and $M_0$ is the magnetization per site.}
 \label{fig:chiral-dynamics}
\end{figure*}

\subsection{Occupied modes and directional motion}

In momentum space, occupied and empty spin-reversal modes meet at two boundaries. These are not physical ends of the periodic chain. Figure~\ref{fig:chiral-dynamics} connects rearrangements near each boundary to excitations moving in opposite directions with sharply different energy costs.

Figure~\ref{fig:chiral-dynamics}(a) shows how the occupied and empty modes form two occupation boundaries. Each dot represents a possible delocalized spin-reversal mode. Blue dots are occupied; open gray dots are empty. Equation~\eqref{eq:momentum-convention} places the empty modes at $-2\pi<p<-\pi$ and the filled modes at $-\pi<p<0$. The middle of the strip is therefore an occupation boundary, which we denote by F, at $x=0$ and $p=-\pi$. Since momenta differing by $2\pi$ are equivalent, the two ends of the strip meet. Together they form a second boundary, denoted by R, approached as $x\to\pm\infty$. F is the ordinary boundary at finite rapidity; R is approached at infinite rapidity. Both separate occupied from empty modes.

A low-energy rearrangement removes a wave from an occupied mode and fills an empty mode across either boundary. The added particle and the hole left behind preserve total spin. Here ``particle--hole pair'' means these two changes of occupation; neither constituent is a bound group of reversed spins. Across F, the transfer goes from just above $p=-\pi$ to just below it and gives a small negative physical momentum. The black curve in Fig.~\ref{fig:chiral-dynamics}(b) shows the minimum energy of one such pair. Close to zero momentum it approaches the dotted straight line. Its finite slope means that a long-wavelength disturbance propagates with a finite speed. Dividing the dressed energy slope $\eps_0'(0)=-\pi\Omega/4$ by the momentum slope $2\pi\rho_0(0)$ gives
\begin{equation}
 \frac{4v_L}{\Omega}=-\frac{1}{2\rho_0(0)}\simeq-1.7940 .
 \label{eq:velocities-main}
\end{equation}
Its energy is therefore $|v_L||\Delta P|$ to leading order. The same density value fixes this velocity and the XXX current in Eq.~\eqref{eq:chirality-density}.

At R, let us remove an occupied mode near the right end of the strip or add one near the left end. These are the orange hole and green particle in Fig.~\ref{fig:chiral-dynamics}(a). Removing a root at $x_h>0$ raises total spin by one; adding one at $-X<0$ lowers it. Both processes have positive physical momentum:
\begin{equation}
 q_h=2\pi\int_{x_h}^{\infty}\rho_0(y)\,dy,\qquad
 q_p=2\pi\int_{-\infty}^{-X}\rho_0^{\rm t}(y)\,dy .
 \label{eq:remote-momenta}
\end{equation}
Here $\rho_0^{\rm t}(x)=a_1(x)-\int_0^\infty a_2(x-y)\rho_0(y)\,dy$ counts available modes on the empty side. The integrals count how many modes lie between the chosen excitation and the remote boundary. They convert the mode coordinate into physical momentum, with scattering already included. The hole and particle energies are $\omega_+=-\eps_0(x_h)$ and $\omega_-=\eps_0(-X)$.

The crucial feature is that energy and momentum vanish at different rates. The density tails give $q_h\simeq2M_0/x_h$ and $q_p\simeq4M_0/X$, while the energy decreases as $\omega\simeq\Omega\ee^{-\pi|x|/2}$. Moving toward R therefore reduces the energy exponentially while reducing momentum only inversely with the mode coordinate. Eliminating that coordinate gives the exponential laws in Eq.~\eqref{eq:headline-essential}. The empty side has twice the leading mode density of the occupied side. This difference explains why adding a particle and making a hole have different exponential scales. The smaller corrections to the densities fix the prefactors and $\kappa_0\simeq1.0373$.

We calculate these energies by solving Eq.~\eqref{eq:rho0-main}, converting the mode labels to physical momenta, and evaluating the prescribed dressed energies. Figure~\ref{fig:chiral-dynamics}(b) makes the directional contrast visible. Its horizontal axis is the physical momentum change $\Delta P$, and its vertical axis is the excitation energy in units of $\Omega/4$. The green and orange curves show the remote particle and hole. Both become almost indistinguishable from zero as positive momentum decreases, while the black curve retains its finite slope on the negative side. The red curve follows the least expensive single particle--hole pair at R. We obtain it by minimizing the sum of the two constituent energies at fixed $q=q_h+q_p$. At the minimum, a small transfer of momentum between the constituents leaves their total energy unchanged to first order: their energy slopes, or group velocities, are equal. This divides the momentum as $q_h/q\to1/3$ and $q_p/q\to2/3$. The first correction to this division gives the $q^{1/6}$ factor in Eq.~\eqref{eq:pair-threshold}. The pair can therefore carry a prescribed small momentum for less energy than a single particle or hole carrying that whole momentum.

The very small energies near the origin of Fig.~\ref{fig:chiral-dynamics}(b) become distinguishable in Fig.~\ref{fig:chiral-dynamics}(c). Let us take the logarithm of the inverse scaled energy and multiply it by momentum, giving $q\ln(\Omega/\omega)$. A power-law energy would make this quantity approach zero. An energy dominated by an exponential in $-1/q$ instead gives a nonzero limit. The orange, green, and red curves approach $\pi M_0$, $2\pi M_0$, and $3\pi M_0$, respectively. These limits identify the three exponential scales directly. A higher curve here means a smaller energy at the same momentum, so the red pair curve corresponds to the softest of the three processes. The short dashed curves include the elementary prefactors in Eq.~\eqref{eq:headline-essential} and show how the particle and hole approach their limits.

Sharing momentum among many remote pairs reduces the energy further. After the infinite-chain limit at fixed pair number $K$, optimized pairs sharing momentum $Q$ cost $K\omega_0(Q/K)$, which tends to zero as $K\to\infty$. Choosing $Q=2\pi-q$ also reaches negative physical momentum $-q$. Allowing arbitrarily many pairs therefore brings the additive spectrum arbitrarily close to zero energy at either momentum sign. Figure~\ref{fig:chiral-dynamics} resolves the distinct single-particle, single-hole, and single-pair energy scales within this larger spectrum. The intensity with which a probe creates each process determines its visibility.

Spin symmetry helps identify these low-energy processes. The generators of the broken rotations obey $[S^x_{\rm tot},S^y_{\rm tot}]=\ii S^z_{\rm tot}$, where $S^a_{\rm tot}=\sum_mS_m^a$. The ordered moment makes the expectation value on the right grow with system size. The two transverse rotations are consequently coupled, as in a ferromagnet~\cite{NielsenChadha,WatanabeMurayama,Hidaka,Leutwyler}. Adding a remote particle lowers the total spin. Applying the total spin-lowering operator instead changes its projection while preserving its magnitude. It produces another state of the same energy within the spin multiplet, called a \emph{descendant}.

We can therefore compare two processes that both lower the spin projection. At $+q$, adding a particle costs $\omega_-(q)$. At $-q$, combining a descendant with a finite-boundary particle--hole pair gives $E_{\rm F}^{(-)}(q)=|v_L|q+o(q)$. Their energy ratio is
\begin{equation}
 \frac{\omega_-(q)}{E_{\rm F}^{(-)}(q)}
 \simeq \frac{\omega_-(q)}{|v_L|q}\longrightarrow0 .
 \label{eq:nonreciprocity}
\end{equation}
The ratio vanishes faster than every power of $q$. The reference XXX Hamiltonian assigns the same remote hole and particle the different costs $+J_{\rm X}q^2/(2M_0)$ and $-J_{\rm X}q^2/(4M_0)$: the selected state is excited in that reference spectrum.

\subsection{What a local spin probe sees}

Excitation energies tell us which spin rearrangements are possible. To determine their visibility in a measurement, we also need their intensities: how strongly the probe couples to each rearrangement. Let us choose the $z$ axis along the ordered moment and compare probes across and along this axis. We use $S_q^a=L^{-1/2}\sum_m\ee^{\ii qm}S_m^a$, for $a=-,z$. This sum combines local spin operators with a spatial phase that selects momentum $q$. The transverse operator $S_q^-$ lowers the total spin projection by one. Starting from the ground state $|0\rangle$, with energy $E_0$, total spin $S$, and maximal projection $S^z_{\rm tot}=S$, it produces the spectrum
\begin{equation}
 S^{-+}(q,\omega)=\sum_\nu|\langle\nu|S_q^-|0\rangle|^2
 \delta\big(\omega-E_\nu+E_0\big).
 \label{eq:response-main}
\end{equation}
Here $|\nu\rangle$ is a final state of energy $E_\nu$. Each squared matrix element measures the strength of one transition. The Dirac delta places that strength at the required excitation energy. Replacing $S_q^-$ by $S_q^z$ gives the longitudinal spectrum $S^{zz}(q,\omega)$: fluctuations along the moment that preserve spin projection.

\begin{figure*}[t]
 \centering
 \includegraphics[width=\textwidth]{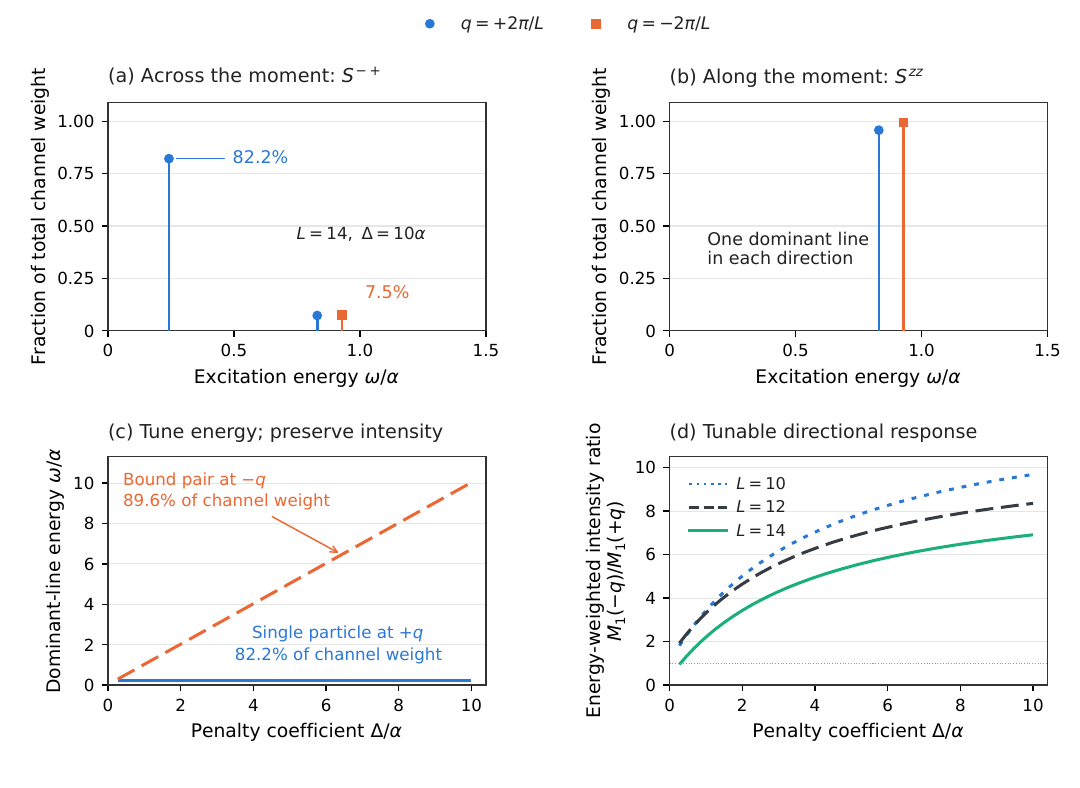}
 \caption{Spin probes separate directional motion, spin symmetry, and bound-state energy. Panels (a)--(c) use $L=14$ and $q=2\pi/L$. Blue and orange denote $+q$ and $-q$. (a) Transverse and (b) longitudinal response at $\Delta=10\alpha$. Positions give excitation energies and heights give fractions of the total intensity for that probe and momentum, integrated over all energies. The displayed window is $0<\omega\le1.5\alpha$. The dominant $-q$ transverse line lies at $10.0285\alpha$, outside this window; the small visible orange line belongs to a different total-spin multiplet. The longitudinal probe resolves a strong low-energy line at each momentum sign. (c) Increasing the bound-group penalty raises the dominant $-q$ pair energy without changing its $89.6\%$ intensity fraction. The $+q$ unbound line keeps its energy and $82.2\%$ fraction. (d) Ratio $M_1(-q)/M_1(+q)$ for $L=10,12,14$. The first moment $M_1$ sums each transition intensity multiplied by its energy over the full spectrum. Its growth shows that the penalty increases the energy-weighted directional contrast. Curves in (c) and (d) start at $\Delta=0.27\alpha$, within the common verified interval where the selected ground multiplets are unchanged.}
 \label{fig:spin-response}
\end{figure*}

We calculate these quantities on periodic chains, with the last spin connected to the first. For each chain, the common eigenstates of the commuting spin operators supply the full energies of $H_+$ and the corresponding wave functions. Acting with a probe on the ground state and projecting the result onto these final states gives the transition intensities. Their sum agrees with the squared norm of the state created by the probe to numerical precision, checking that the response has been accounted for. Figure~\ref{fig:spin-response} shows the discrete spectral lines of these finite systems; its first three panels use $L=14$ and the smallest nonzero momentum magnitude, $q=2\pi/L$.

Figure~\ref{fig:spin-response}(a) compares the transverse response in the two momentum directions at $\Delta=10\alpha$. The horizontal position of a line gives its energy; its height gives its share of the total intensity at that momentum. The blue line at $+q$ carries $82.2\%$ of the intensity at an energy $0.2420\alpha$. It adds an unbound mode. The strongest transition at $-q$ instead creates a state containing a bound pair. This orange line carries $89.6\%$ at $10.0285\alpha$, above the displayed low-energy window. The visible orange line near $0.9279\alpha$ carries only $7.5\%$. Thus the weak low-energy response at $-q$ accompanies a strong response at higher energy. The line heights use the total intensity over \emph{all} energies, so the displayed heights need not sum to one. The same contrast occurs for $L=10,12$: $82$--$84\%$ of the positive-momentum weight lies in an unbound excitation, while about $90\%$ of the negative-momentum weight lies in a bound-pair state. The probe adds one net spin reversal while also rearranging those already present. A final state with two bound reversals and one fewer unbound reversal has precisely this net change.

Figure~\ref{fig:spin-response}(b) shows what changes when the probe acts along the moment. At $L=14$, one low-energy line carries $95.7\%$ of the longitudinal weight at $+q$ and another carries $99.3\%$ at $-q$. Their energies differ even though their total longitudinal weights, summed over all energies, are equal. Spin symmetry explains why these are different from the dominant transverse lines. The latter have final total spin $S-1$. A longitudinal probe preserves projection $S$, so it can reach only total spin $S$ or $S+1$. For a final spin-$S$ multiplet, its spin-lowered partner has exactly $2/S$ times the \emph{absolute} longitudinal intensity in the transverse response. The separate normalizations in panels (a) and (b) then give the smaller transverse peaks at the same energies as the corresponding longitudinal peaks. Across the sampled chains, the strongest negative-momentum longitudinal line carries $94$--$100\%$ of its channel weight and has energy near $|v_L|q$, connecting it to the finite-velocity branch.

Figure~\ref{fig:spin-response}(c) exposes a useful control of the response. Increasing $\Delta$ changes the energy of a bound group while preserving its wave function. Throughout the common verified interval $\Delta\ge0.27\alpha$ for the calculated chains, the ground multiplet also remains fixed. The dominant bound-pair energy at $L=14$ therefore rises as $0.0285\alpha+\Delta$, while its intensity stays at $89.6\%$. The positive-momentum unbound line stays at $0.2420\alpha$ with $82.15\%$ of its channel weight. The penalty thus adjusts the energy needed to excite the bound pair independently of how strongly the probe excites it.

Finally, Fig.~\ref{fig:spin-response}(d) combines line positions and intensities through the first frequency moment, $M_1(q)=\int_0^\infty\omega S^{-+}(q,\omega)\,d\omega$. This is the sum of each transition's intensity multiplied by its energy, including lines outside the window of panel (a). The ratio $M_1(-q)/M_1(+q)$ grows with the penalty for all three displayed sizes and reaches $6.91$ at $L=14$ and $\Delta=10\alpha$. Direction, probe orientation, and bound-state energy therefore provide distinct ways to resolve the collective spin motion. These finite-chain results specify concrete spectroscopic signatures of the constructed interaction.

\section{Magnetic response and low-temperature entropy}
\label{sec:soft}

\begin{figure*}[t]
 \centering
 \includegraphics[width=0.94\textwidth]{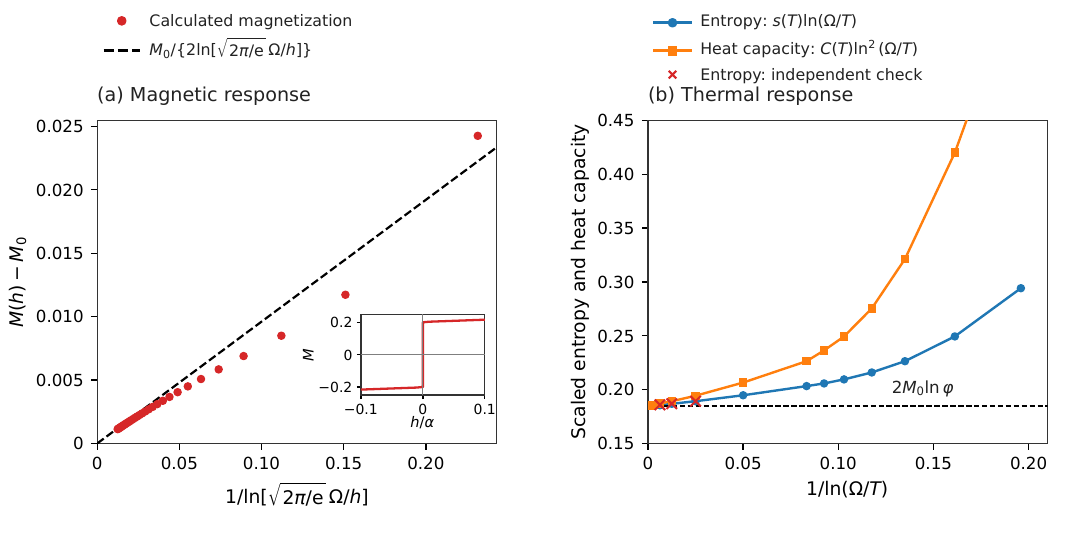}
 \caption{Magnetic and thermal responses of the selector. (a) The magnetization increase $M(h)-M_0$ approaches $M_0/[2\ln(\sqrt{2\pi/\ee}\,\Omega/h)]$, where $h$ is the uniform field and $\Omega=\pi^2\alpha/2$. Deviations reveal subleading logarithmic corrections. The inset shows both symmetry-selected branches of the magnetization per site $M(h)$. (b) Entropy density $s(T)$ and specific heat per site $C(T)=T\,ds/dT$, multiplied respectively by $\ln(\Omega/T)$ and $[\ln(\Omega/T)]^2$, approach $2M_0\ln\varphi$. Here $T$ is temperature, $\Omega=\pi^2\alpha/2$, and $\varphi=(1+\sqrt5)/2$ is the golden ratio. Crosses show the independently calculated entropy. The entropy coefficient is controlled for the full additive parent; the heat-capacity curve tests the reduced real-root equations.}
 \label{fig:soft}
\end{figure*}

Let us apply a weak field along the ordered moment. A field $h>0$ adds $-hS^z_{\rm tot}$ and favors removing the least expensive spin-reversal modes, which lie at large positive rapidity. The occupied boundary $X(h)$ moves outward logarithmically as the field decreases. Counting the removed tail and the scattering-induced readjustment gives Eq.~\eqref{eq:headline-field}. This readjustment fixes the scale $\sqrt{2\pi/\ee}\,\Omega=7.5026\alpha$ inside the logarithm. The moment stays finite while the susceptibility diverges [Fig.~\ref{fig:soft}(a)].

Equivalently, the minimum energy density $e(m)$ at fixed magnetization $m>M_0$ obeys
$\ln\{[e(m)-e(M_0)]/\alpha\}=-M_0/[2(m-M_0)]+O[\ln(1/(m-M_0))]$,
so a small increase above $M_0$ costs exponentially little energy. Within $|m|\le M_0$, the degenerate spin multiplet makes this energy flat. Complete field scans select all-real multiplets through $L=14$. For chains through $L=256$, an enumerated family of real-root trial states gives magnetization staircases whose selected-state energies remain closely aligned after the exact-operator correction.

Temperature accesses modes with energies comparable to or below $T$. For $T\ll\Omega$, these lie beyond the rapidity scale
\begin{equation}
 X_T=\frac2\pi\ln\frac{\Omega}{T}+O(1).
 \label{eq:thermal-edge-main}
\end{equation}
The number of such modes per site is of order $1/X_T$. Their entropy is therefore proportional to system size, with a density that decreases only logarithmically on cooling. Here we take $L\to\infty$ before lowering $T$. A single spin multiplet on a finite chain contributes $O(\ln L)$ total entropy and hence vanishing entropy per site.

Let us count these configurations while including the scattering-induced change in available modes. This gives
\begin{align}
 s(T)&=\frac{A}{\ln(\Omega/T)}
       +O\!\left(\frac{\ln\ln(\Omega/T)}{[\ln(\Omega/T)]^2}\right),\nonumber\\
 C(T)&=\frac{A}{[\ln(\Omega/T)]^2}
       +O\!\left(\frac{\ln\ln(\Omega/T)}{[\ln(\Omega/T)]^3}\right),\nonumber\\
 A&=2M_0\ln\varphi\simeq0.1849 .
 \label{eq:thermal-log-laws}
\end{align}
Here $C(T)=T\,ds/dT$ is the specific heat per site and $\varphi$ is the golden ratio. In the thermally active tails, the occupied fraction approaches $\varphi^{-2}$. Both tails contribute to $A$, while the linear edge supplies a smaller term proportional to $T$. The golden ratio thus enters through the counting of interacting modes.

For $\Delta>\Omega/2$, all bound strings together change the additive free energy by at most $2T\ee^{-(\Delta-\Omega/2)/T}/(1-\ee^{-\Delta/T})$. This uniform bound fixes the leading entropy coefficient for the full string problem. Numerical solutions also support the specific-heat prediction of the smooth reduced equations [Fig.~\ref{fig:soft}(b)]. The spin-rotation-invariant Gibbs ensemble counts all accessible orientations and excitations.

\section{Added exchange reconstructs the remote edge}
\label{sec:exchange}
\label{sec:Mtransition}

Let us test how the activated spectrum responds to a familiar interaction. Adding Heisenberg exchange gives
\begin{equation}
 H(J)=H_++JQ_2,\qquad j=J/\alpha .
 \label{eq:Hj}
\end{equation}
Changing $J$ shifts the energies of the common eigenvectors and can therefore change which one is the ground state. For $\Delta=10\alpha$ and $|j|\le2$, the dressed bound-group energies obey $\eps_{n\ge2}\ge\lfloor n/2\rfloor\Delta-\Omega/2-\max(J,0)>5.53\alpha$. The positive gap keeps these groups out of the additive ground state throughout this interval, so its occupation is determined by real roots.

The outer boundaries are
\begin{equation}
 j_{\rm F}=-1.1340,\qquad j_{\rm AF}=1.1932 .
 \label{eq:boundaries-main}
\end{equation}
Below $j_{\rm F}$ the selected state is fully aligned; above $j_{\rm AF}$ it is the antiferromagnetic reference ground-state vector. At each boundary a filled or empty interval opens, with energy proportional to momentum squared. This is a one-dimensional Lifshitz transition, with dynamical exponent $z=2$ [Fig.~\ref{fig:exchange}]. Between the outer boundaries, negative exchange selects one finite occupied interval; positive exchange adds an occupied tail from negative infinity. Both signs restore quadratic remote excitations.

Near $j=0$, even weak exchange competes with the exponentially small selector energy. Its contribution decreases only as inverse rapidity squared, so the remote boundary moves as $\ln(1/|j|)$. Removing remote roots for $j<0$, or filling the opposite tail for $j>0$, gives the continuous magnetization with divergent slope in Eq.~\eqref{eq:M-main}. The ground-state energy has a continuous first derivative and a divergent second derivative.

The crossover momentum is $q_{\rm cr}\sim\pi M_0/\ln(1/|j|)$. Above this scale the activated spectrum remains visible; below it the energy becomes quadratic. This crossover quantifies how precisely exchange must be tuned to resolve the activated behavior at a chosen wavelength.

\section{Spatial range and sensitivity to smoothing}
\label{sec:locality}

\begin{figure*}[t]
 \centering
 \includegraphics[width=\textwidth]{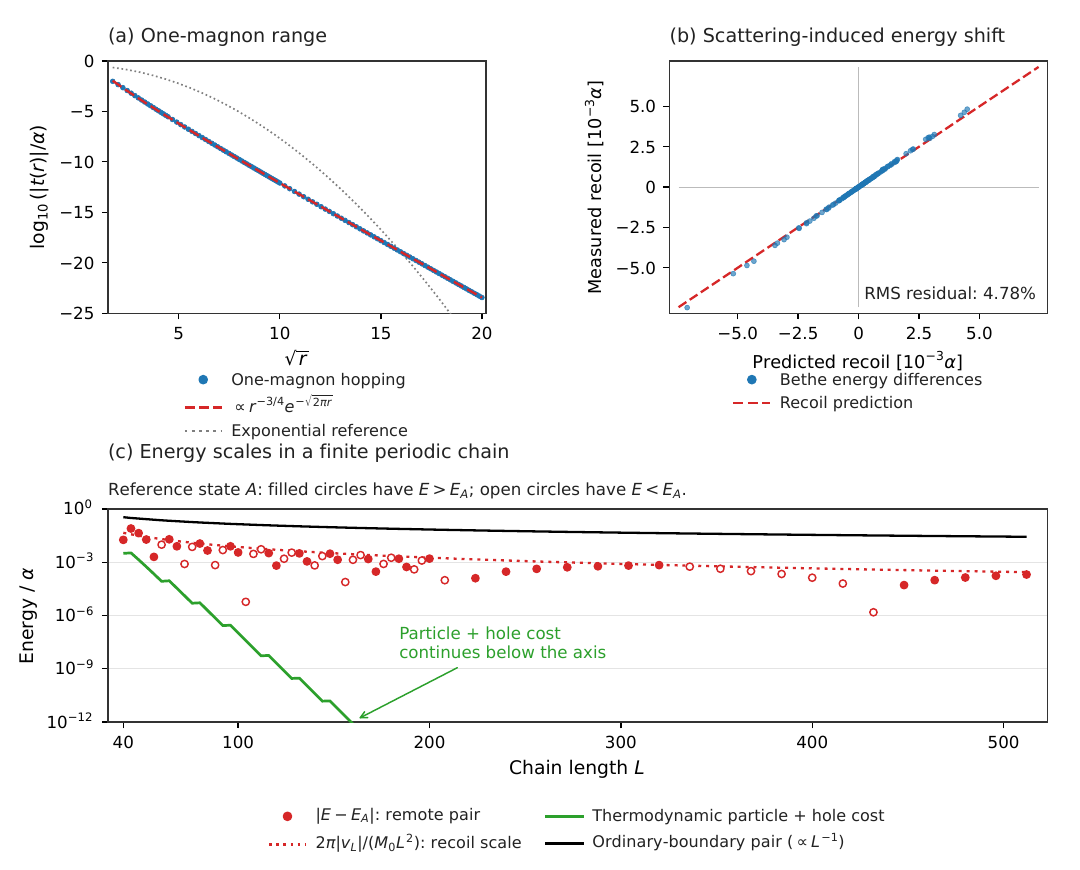}
 \caption{Spatial range, scattering recoil, and energy scales of finite periodic chains. (a) One-magnon hopping magnitudes follow Eq.~\eqref{eq:stretched}; the dotted gray curve is an exponential reference. (b) Recoil is the finite-chain energy change beyond the thermodynamic particle and hole cost. The reference $A$ has $N=\operatorname{round}(N_0L)$ consecutively occupied modes, and $E_A$ is its energy. A hole at $x_h>0$ has cost $\omega_h=-\eps_0(x_h)$; a particle at $-X<0$ has cost $\omega_p=\eps_0(-X)$. The measured recoil $E-E_A-(\omega_h+\omega_p)$ is compared with the linear and quadratic response of the ordinary boundary to their scattering phase. The $135$ pairs have $112\le L\le512$, $x_h>8$, and $X>8$. The root-mean-square residual is $4.78\%$ of the measured recoil scale. (c) The smallest-momentum remote pair, denoted $D(0,0)$, moves the outermost occupied mode to the first empty slot across the periodic momentum boundary. Its finite-chain splitting $|E_{D(0,0)}-E_A|$ is shown in red. The green curve is the thermodynamic cost $\omega_h+\omega_p$ at the same pair rapidities, exactly the sum subtracted in (b). The dotted red curve shows a characteristic $L^{-2}$ recoil scale; the black curve is a distinct pair near the ordinary boundary with energy proportional to $L^{-1}$. Filled red circles indicate $E_{D(0,0)}>E_A$, and open red circles indicate $E_{D(0,0)}<E_A$. The green curve continues below the displayed range. The red points compare each rearranged state with the specified reference $A$; the dotted curve shows the characteristic recoil scale.}
 \label{fig:locality-ring}
\end{figure*}

Let us translate the spectral construction into motion along the chain. The one-magnon hopping amplitude $t(r)$ moves a reversed spin by $r$ sites through an aligned background. At large separation,
$|t(r)|\sim(\pi^{7/4}/4)\alpha r^{-3/4}\ee^{-\sqrt{2\pi r}}$,
including the analytically determined prefactor. Numerically evaluated amplitudes follow this law over twenty-one decades [Fig.~\ref{fig:locality-ring}(a)]. The exponential depends on $\sqrt r$ rather than $r$, giving a \emph{stretched exponential}. It decreases faster than every power but has no finite exponential decay length~\cite{KatznelsonHarmonic}. The interaction therefore extends to all separations with rapidly decreasing strength.

With two reversed spins, we can separate independent hopping from their correlated motion. On periodic chains with $L=12,14,16$, the largest matrix elements of the correlated part fall by approximately a factor of two for each added site of initial separation over the resolved range. Both individual and correlated motion therefore have rapidly decreasing amplitudes in the sectors studied.

We can also test the sharp cutoff in the weights defining the nonlocal charge. The original construction includes only negative rapidities. Here, \emph{smoothing} means replacing this abrupt cutoff by the gradual weight $F_\delta(x)=1/(1+\ee^{x/\delta})$, where $\delta>0$ sets the transition width in the mode label $x$. This changes the interaction itself and restores an energy tail that decreases as a power, with amplitude proportional to $\delta^2$. Occupation then ends at $X_\delta\simeq(4/\pi)\ln(1/\delta)$, producing a quadratic remote mode. This boundary fixes the momentum and energy window in which the activated behavior survives. Exchange and smoothing thus provide two quantitative controls over the observable spectrum.

\section{Scattering-induced energy shifts in finite periodic chains}
\label{sec:ring}

Consider a finite periodic chain, whose last site is connected to the first. A spin-reversal wave must return to the same phase after traveling around the chain, including the phases acquired when it scatters from the other waves. These conditions determine the allowed momenta together. Creating a particle or a hole near the very soft occupation boundary therefore shifts other occupied modes, including those near the ordinary boundary with finite velocity. Their additional finite-size energy contribution is \emph{recoil}.

Let us specify the two states being compared. The reference state $A$ fills $N=\operatorname{round}(N_0L)$ consecutive allowed modes, where $N_0\simeq0.3079$ is the occupied fraction of the infinite-chain state. This choice provides a finite-chain representative of the selected half-line occupation. A particle--hole rearrangement removes one occupied mode and fills an empty one, preserving $N$. Let $x_j^{(A)}$ and $x_j$ denote the roots of the reference and rearranged states, respectively. Both sets satisfy the finite-chain Bethe equations. We evaluate their total energies, $E_A$ and $E$, using the additive assignment $\sum_jw_1(x_j)$, so that
\[
 E-E_A=\sum_{j=1}^{N}\bigl[w_1(x_j)-w_1(x_j^{(A)})\bigr].
\]
The difference includes the changes in every root, not just the particle and hole.

A positive $E-E_A$ means that the rearranged state lies above $A$; a negative value means that it lies below $A$. The reference is fixed by its occupation block rather than by minimizing the energy over all finite-chain states. These differences therefore measure the ordering of two specified levels. By contrast, an excitation gap is measured from the minimum energy of the finite system and is nonnegative. A negative difference here identifies a lower-energy finite-chain arrangement relative to $A$, consistent with finite-size corrections that vanish in the thermodynamic limit.

For a remote pair, the positive thermodynamic cost is $\omega_+(q_h)+\omega_-(q_p)$. Recoil is the remaining contribution $E-E_A-[\omega_+(q_h)+\omega_-(q_p)]$. It describes the effect of discrete allowed momenta beyond the collective readjustment already included in the dressed energies. The scattering phase $\theta_2(x)=2\arctan(x/2)$ approaches $\pm\pi$ with correction $-4/x+O(x^{-3})$. Thus the excitation energy becomes exponentially small at large rapidity, whereas the phase shift it produces decreases only algebraically. A very inexpensive rearrangement can consequently produce an appreciable energy shift near the ordinary boundary.

Finite-size Bethe-ansatz methods quantify this response~\cite{deVegaWoynarovich,IzerginKorepinReshetikhin}. Let $E_A(\zeta)$ be the reference energy continued to a small common shift $\zeta$ of its mode labels, with the roots recalculated and $E_A(0)=E_A$. The derivative $E_A'(0)$ measures the first-order energy response. We write $E_A'(0)=(\pi|v_L|/L)\zeta_L$, where the dimensionless offset $\zeta_L$ records the placement of the ordinary occupation boundary among the discrete modes. The leading recoil contribution $E_R$ for the smallest-momentum remote pair is then
\begin{equation}
 E_R\simeq-\frac{2\pi|v_L|}{M_0L^2}\,\zeta_L .
 \label{eq:recoil-gap}
\end{equation}
Its scale is generically $L^{-2}$, while its sign depends on the offset. For $135$ remote pairs with $112\le L\le512$, the predicted linear and quadratic responses reproduce recoil with a root-mean-square residual of $4.78\%$ [Fig.~\ref{fig:locality-ring}(b)]. Exact-operator corrections are much smaller in these families; for a remote particle they fall from $2.7\times10^{-5}\alpha$ at $L=48$ to $1.4\times10^{-12}\alpha$ at $L=256$.

Figure~\ref{fig:locality-ring}(c) shows how strongly the two energy scales can differ. At $L=112$, the thermodynamic pair cost is $5.42\times10^{-9}\alpha$, while $E-E_A\simeq-0.0054\alpha$. The negative value means that the rearranged state lies below the chosen reference because the collective finite-size shift outweighs the positive pair cost. The panel displays $|E-E_A|$ on a logarithmic axis; filled and open red circles distinguish states above and below $A$.

The physical consequence is that an exceptionally soft excitation still interacts with the rest of the magnet. Its finite-chain energy difference can be dominated by the response of modes with finite velocity. An algebraic dependence on chain length can therefore coexist with an activated infinite-chain dispersion. To separate these contributions, let us increase $L$ at fixed physical momentum $q>0$, for which recoil decreases as $q/L$, and only then reduce $q$. Following the smallest allowed pair momentum instead makes $q$ decrease as $1/L$: the activated cost becomes exponentially small in $L$, while recoil remains algebraic. This distinction gives a practical strategy for numerical spectra: compare states at fixed momentum across increasing chain lengths, account for recoil, and then examine the long-wavelength energy law. The finite-size shifts thus reveal the interaction between the two excitation sectors and explain how the directional thermodynamic spectrum emerges from discrete levels.

\section{Conclusions}
\label{sec:conclusion}

We have shown how an ordered quantum state can sustain sharply different energy scales for collective motion in opposite directions. By assigning new energies to common Heisenberg eigenstates, we construct a nonlocal selector whose thermodynamic ground state has spontaneous magnetization, zero occupation entropy per site, and a persistent current of the reference Heisenberg Hamiltonian. The construction separates two physical questions: how spins are organized in a wave function, and how much energy their collective rearrangement requires.

At the maximally chiral point, one elementary branch retains a finite propagation speed, while the opposite branches have energies that decrease exponentially with inverse momentum. Their origin is a mismatch between energy and state counting: the energies fall exponentially along the sequence of available modes, while the mode density has an algebraic tail. The same mechanism makes the ordered moment unusually easy to change. It produces a divergent weak-field susceptibility and, in the additive thermodynamic description with the stated bound-group penalty, an entropy per spin that vanishes only inversely with a logarithm of temperature. Its leading coefficient contains the golden ratio. Thus magnetic order coexists with exceptionally inexpensive rearrangements and a slowly depleted set of thermally accessible states.

The spin response makes this structure experimentally distinguishable. Exact calculations on periodic chains identify a strong unbound excitation at one momentum sign and a strong bound-pair excitation at the other. Changing the bound-group penalty shifts the latter's energy while preserving its intensity throughout the interval with the same selected ground state. Spin-rotation symmetry supplies exact relations between corresponding longitudinal and transverse intensities. Together, these results show how the wave functions determine the strength of a signal while the commuting interactions control its energy.

The microscopic analysis also identifies the spatial and size scales behind these effects. The one-spin hopping amplitude has a stretched-exponential tail. Added exchange and smoothing restore quadratic excitations at the lowest momenta and determine the range in which the activated spectrum remains visible. Analytic correspondence bounds and exact finite-chain calculations provide complementary tests of the spin operator and its thermodynamic description. The calculated recoil explains how scattering shifts finite-chain energies and gives a procedure for resolving the thermodynamic branches by increasing the chain length at fixed momentum.

For an engineered realization governed by the selector interaction, we propose three complementary probes. Momentum-resolved spin spectroscopy, using a weak magnetic modulation with controlled traveling direction, frequency, and wavelength, would compare energies and intensities at opposite momenta. Modulation parallel to the ordered moment probes the longitudinal response. Circularly polarized transverse modulation of the appropriate handedness selects the spin-lowering response calculated here. A particularly direct test would follow the bound-pair line as its energy changes while its integrated intensity stays fixed. Weak-field magnetometry would test the inverse-logarithmic approach to a finite spontaneous moment. Low-temperature calorimetry would reconstruct the entropy by integrating the heat capacity divided by temperature. This would test both the slow loss of accessible states on cooling and its predicted coefficient. Varying the added exchange would provide an additional check through the crossover to quadratic motion. These linked signatures turn the designed spectrum into a set of independently testable magnetic, dynamical, and thermal predictions. They provide benchmarks for quantum simulation and a route toward other interacting states with prescribed collective dynamics.

\begin{acknowledgments}
The author thanks Fabian Essler, Boris Svistunov, and Nikolay Prokofiev for illuminating discussions. This work was supported by the Armenian Higher Education and Science Committee under the ARPI Remote Laboratory grant 24RL-1C024.
\end{acknowledgments}

\end{document}